\documentclass[british,superscriptaddress,amsmath,amssymb,preprint,pre]{revtex4-1}

\usepackage{lmodern}
\usepackage{lmodern}
\usepackage[T1]{fontenc}
\usepackage[utf8]{inputenc}
\usepackage{float}
\usepackage{amsmath}
\usepackage{graphicx}

\makeatletter

\usepackage{microtype}
\usepackage{chngcntr}
\counterwithout{figure}{section}

\usepackage[font=it]{caption}
\usepackage{xr}
\usepackage{siunitx}
  \DeclareSIUnit\Molar{M}
  \DeclareSIUnit\min{min}

\usepackage[bottom]{footmisc}
\usepackage{babel}
\makeatother

\usepackage{babel}
\begin{document}
\global\long\def\rcqd{\mbox{\ensuremath{\mathrm{BRC_{QD}}}}}
 \global\long\def\rcox{\mbox{\ensuremath{\mathrm{BRC_{OX}}}}}
 \global\long\def\icm{\mbox{\ensuremath{{\rm cm}^{-1}}}}
 \global\long\def\pdw{P_{y^{-}}}
 \global\long\def\pup{P_{y^{+}}}

\title{Quantum Coherence as a Witness of Vibronically Hot Energy Transfer
in Bacterial Reaction Centre}

\author{David Paleček}\email{Current address: Department of Chemistry, University of Zurich, Winterthurerstrasse 190, CH-8037, Zurich}
\affiliation{Department of Chemical Physics, Lund University, P.O. Box 124, SE-22100 Lund, Sweden}

\affiliation{Department of Chemical Physics, Charles University in Prague, Ke
Karlovu 3, 121 16 Praha 2, Czech Republic}

\author{Petra Edlund}
\affiliation{Department of Chemistry and Molecular Biology, University of Gothenburg, Box 462, SE-40530 Gothenburg, Sweden}

\author{Sebastian Westenhoff}
\affiliation{Department of Chemistry and Molecular Biology, University of Gothenburg,
Box 462, SE-40530 Gothenburg, Sweden}

\author{Donatas Zigmantas}\email{donatas.zigmantas@chemphys.lu.se}
\affiliation{Department of Chemical Physics, Lund University, P.O. Box 124, SE-22100 Lund, Sweden}

\begin{abstract}
Photosynthetic proteins have evolved over billions of years so as
to undergo optimal energy transfer to the sites of charge separation.
Based on spectroscopically detected quantum coherences, it has been
suggested that this energy transfer is partially wavelike. This conclusion
critically depends on assignment of the coherences to the evolution
of excitonic superpositions. Here we demonstrate for a bacterial reaction
centre protein that long-lived coherent spectroscopic oscillations,
which bear canonical signatures of excitonic superpositions, are essentially
vibrational excited state coherences shifted to the ground state of
the chromophores . We show that appearance of these coherences is
brought about by release of electronic energy during the energy transfer.
Our results establish how energy migrates on vibrationally hot chromophores
in the reaction centre and they call for a re-examination of claims
of quantum energy transfer in photosynthesis. 
\end{abstract}
\maketitle

For efficient photosynthesis, energy migrates through large chromophore
assemblies to the active site of charge generation. This transfer
is generally downhill in energy, but every energy transfer step must
obey the law of energy conservation. This means that vibrational or
environmental modes take up the excess energy of each transfer step.
However, the identification of these critical modes and their coupling
to electronic transitions is difficult, because they are not readily
observable in conjunction with energy transfer using current spectroscopic
methods.

Following absorption of a short laser pulse, the electron cloud and
atoms oscillate coherently across the molecules; electronic and vibrational
coherences are established. Coupling to the environment, for example
to solvent molecules or a protein binding pocket, leads to dephasing
of these coherences across the molecules. Vibrational wave packets
have been observed in photosynthetic proteins using femtosecond transient
absorption experiments in the visible and near infrared spectral regions
\citep{Vos1994,Savikhin1994}. In addition, polarisation-resolved
experiments provided early evidence for electronic coherences in photosynthetic
proteins \citep{Arnett1999,Savikhin1997}. The advent of two-dimensional
electronic spectroscopy (2DES) opened a new avenue for studying electronic
and vibrational coherences \citep{Jonas2003,Engel2007,Nemeth2008}.
The higher dimensionality of the data in these experiments leads to
less spectral congestion, thereby enabling more detailed insight \citep{Brixner2005,Zigmantas2006PNAS,Thyrhaug2016}.

In the 2DES experiment a sequence of three ultrashort laser pulses
(with time delays $t_{1}$ and $t_{2}$ between them) excites the
sample. A third order polarisation is induced in the sample and the
emitted signal field is detected with a fourth pulse at $t_{3}$ after
the last excitation pulse. The evolution of the excited states is
then resolved in the two-dimensional spectra relating excitation ($\omega_{1}$)
and detection frequencies ($\omega_{3}$), which are Fourier transform
conjugates of the $t_{1}$ and $t_{3}$ time delays, respectively
\citep{Jonas2003}. Coherences between states are detected as oscillatory
signals along $t_{2}$. It is very challenging to distinguish between
electronic or vibrational origin of the coherences, but polarisation
resolved 2DES \citep{Hochstrasser2001,Schlau-Cohen2012,Westenhoff2012,Lim2015},
an additional Fourier transform over population time $t_{2}$ \citep{Li2013,Milota2013},
and theoretical analyses \citep{Butkus2012,Seibt2013} have been developed
to this end.

The first reports on 2D electronic spectra of light-harvesting proteins
created excitement, because it was concluded that electronic coherences
live for more than a picosecond \citep{Engel2007,Collini2010,Schlau-Cohen2012}.
This would suggest that energy could migrate wave-like through the
photosynthetic proteins. This far-reaching conclusion relies on the
assignment of the observed oscillations to electronic coherences.
From the outset, electronic coherences surviving in proteins for picoseconds
seem questionable, because the environment as well as intramolecular
vibrations readily induce fluctuation of electronic transitions on
a femtosecond time scale. Thus electronic coherences should decay
much faster than vibrational coherences, which typically dephase within
a couple of picoseconds.

\begin{figure}[h]
\begin{centering}
\includegraphics[width=0.5\columnwidth]{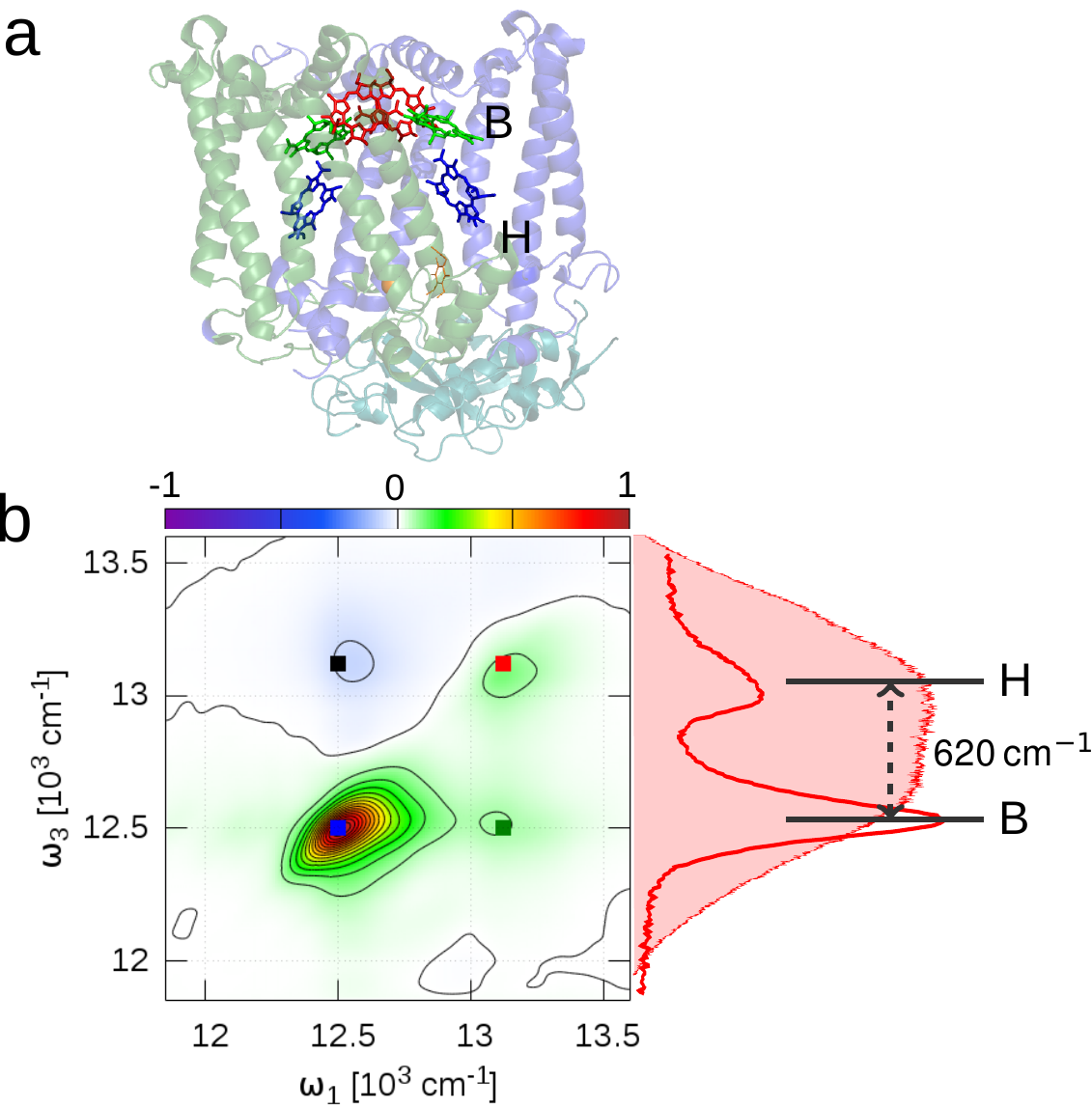} 
\par\end{centering}

\caption{(a) The structure of $\mathrm{RC_{sph}}$. The special pair is coloured
red, the accessory bacteriochlorophylls green and the bacteriopherophytins
blue. (b) The $\mathrm{77\,K}$ 2D spectrum of the oxidised $\mathrm{RC_{sph}}$
at $t_{2}=24\,\mathrm{fs}$, together with the absorption spectrum
and the shaded laser spectrum. The chromophores form a spectroscopic
aggregate and give rise to two excitonic absorption bands, which are
denoted as B, and H. The energy gap between these bands is $\mathrm{\sim620\,cm^{-1}}$.
The absorption of the oxidised special pair $\mathrm{P^{+}}$ is not
observed.}

\label{fig:fig1} 
\end{figure}

The issue of mismatching timescales comes to a head in the reaction
centre protein (Fig.~\ref{fig:fig1}a). There, a particular set of
oscillations exist, which are candidates for excitonic coherences
with the lifetime of picoseconds \citep{Lee2007,Westenhoff2012}.
However, excitations in the reaction centre transfer away from the
participating states within $\mathrm{200\,fs}$ \citep{Jonas1996}.
Since coherences cannot ``outlive'' populations, an unsettling mismatch
between the experimental observation and this fundamental quantum
mechanical principle arises \citep{Ryu2014}.

\subsection*{2DES spectroscopy of $\mathrm{RC_{sph}}$.}

We performed 2DES experiments on \emph{Rhodobacter sphaeroides} reaction
centres grown from the carotenoid deficient R-26 strain ($\mathrm{RC_{sph}}$).
The protein comprises two strongly coupled bacteriochlorophyll\,$a$
molecules, forming a special pair, two accessory bacteriochlorophylls~$a$,
and two bacteriopheophytins~$a$ (Fig.~\ref{fig:fig1}a). To enable
the use of 2DES at a high laser repetition rate ($\mathrm{20\,kHz}$)
and to avoid complication of the spectra from charge transfer signals,
we used $\mathrm{RC_{sph}}$ with a chemically oxidised special pair
($\mathrm{P^{+}}$) at $\mathrm{77\,K}$. This shortens the photocycle
considerably by preventing charge separation, without altering the
rate of energy transfer to the special pair \citep{Jackson1997}.

The linear absorption spectrum of the reaction centre at $\mathrm{77\,K}$
(Fig.~\ref{fig:fig1}b) shows two distinct peaks. The peaks are separated
by $\mathrm{\sim620\,cm^{-1}}$ and are associated with the excitonic
bands B and H, which have dominant bacteriochlorophyll\,$a$ and
bacteriopheophytin~$a$ contributions respectively (Fig.~\ref{fig:fig1}b).
The diagonal peaks in the 2D absorption spectra measured at $t{}_{2}=\mathrm{24\,fs}$
match the absorption bands (Fig.~\ref{fig:fig1}b). The positive
cross-peak below the diagonal indicates coupling and excitation energy
transfer between H and B. The negative-valued region above the diagonal
arises from the excited state absorption of B. This absorption also
masks the upper H-B cross-peak. We did not observe any distinct features
of the two branches of $\mathrm{RC_{sph}}$ in the 2DES data.

\begin{figure}[H]
\begin{centering}
\includegraphics[width=0.5\columnwidth]{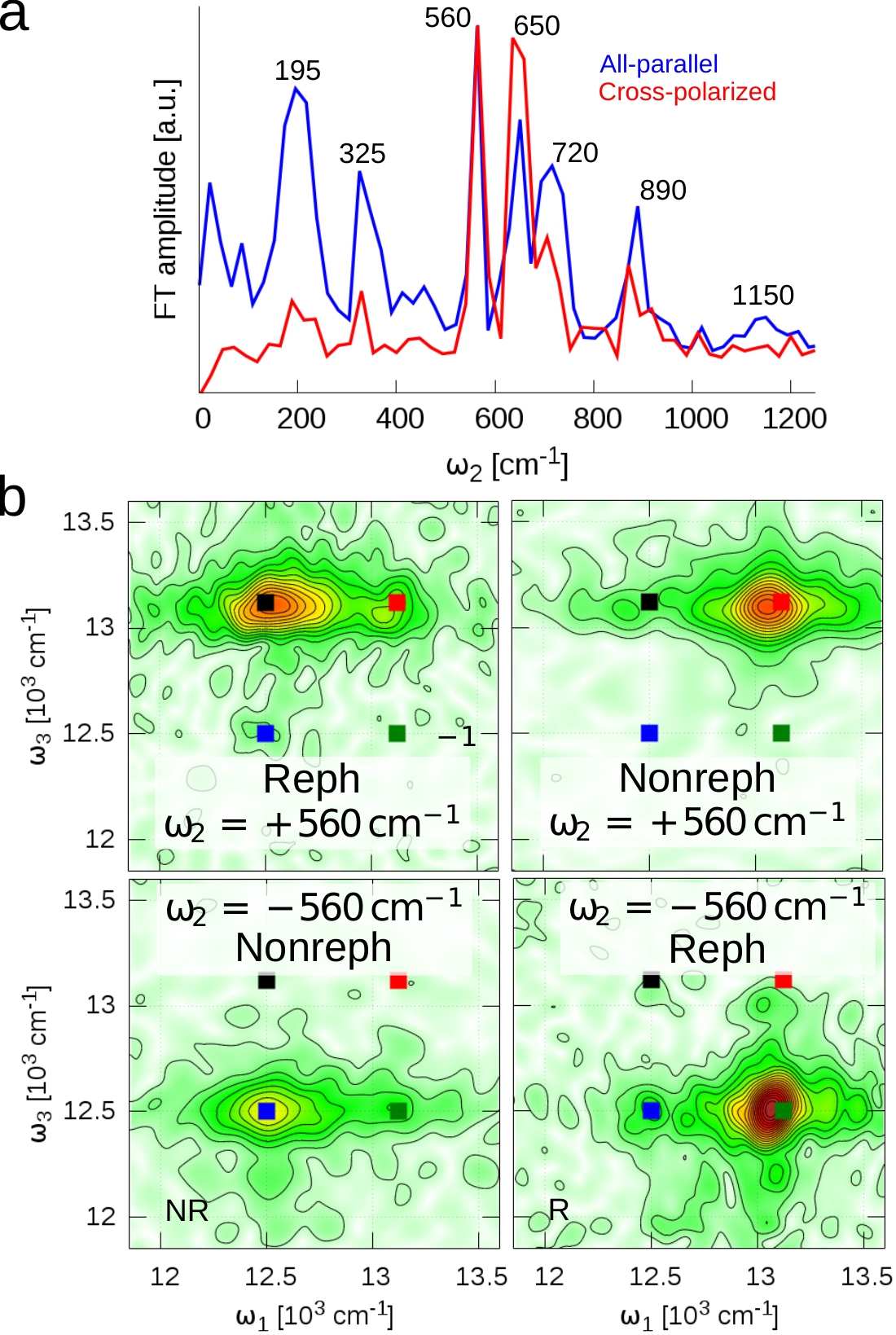} 
\par\end{centering}

\caption{Suppression of the purely vibrational coherences by the polarisation-resolved
2DES. (a) The amplitude of integrated oscillation maps over \textbf{$\omega_{1}$}
and \textbf{$\omega_{3}$} show all the beating frequencies present
in the all-parallel and cross-polarised measurements. The relative
enhancement of the modes with frequency $\omega_{2}=560,\,650\,\mathrm{cm^{-1}}$
close to the excitonic splitting between B and H is clearly visible
in the cross-polarised experiment. The spectra are normalised at $\mathrm{560\,cm^{-1}}$.
(b) \textbf{$\omega_{2}=\pm560\,\protect\icm$} oscillation maps,
extracted from the cross-polarised spectra, measured between $\mathrm{216-1620\,fs}$.
The rephasing part oscillates exclusively off-diagonal and the non-rephasing
part oscillates on diagonal only. The sign of the \textbf{$\omega_{2}$}
frequency indicates the phase evolution direction in the complex plane.
All the features of the presented maps point (misleadingly) to the
electronic origin of the beatings. The coloured rectangles depict
positions of the main peaks in the 2D spectrum.}

\label{fig:fig2} 
\end{figure}

Features in the 2D spectrum decay or rise monotonously along $t_{2}$and
have oscillatory components \citep{Westenhoff2012}. The former reports
on the \emph{populations} of excited states and the latter indicates
\emph{coherences}. To extract the coherent oscillations we subtracted
multiexponential fits in each point of the 2D spectrum. The amplitude
of these oscillatory residuals, integrated across the 2D spectra,
reveal frequencies of 195,~325,~560,~650,~725,~890,~$1150\,\icm$
(blue line, ``all-parallel'' measurement, Fig.~\ref{fig:fig2}a).
All of these frequencies can be identified as peaks in the resonance
Raman spectra of B \citep{Cherepy1997b,Czarnecki1997}.

\subsection*{Picosecond oscillations in $\mathrm{RC_{sph}}$ bear all signatures
of electronic coherences.}

The assignment of the origin of these coherences was achieved in two
steps. Firstly, we used a cross-polarisation scheme (relative polarisations
of $\pi/4,-\pi/4,\pi/2,0$ for laser pulses 1 to 4), which strongly
suppresses signals of purely vibrational coherences (produced by Franck-Condon
excitation of vibrational wavepackets) \citep{Westenhoff2012,Schlau-Cohen2012}.
Indeed, the measurement shows that the amplitudes of the frequencies
at 195,~325,~725,~890 and $\mathrm{1150\,cm^{-1}}$ are decreased
relative to the frequencies at $560,\,650\,\mathrm{cm^{-1}}$ (Fig.~\ref{fig:fig2}A,
red line, ``cross-polarised''). The sustained frequencies are close
to the energy gap between B and H ($\mathrm{620\,cm^{-1}}$, see Fig.~\ref{fig:fig1}b).
This has previously been considered as an indication for an electronic
contribution to the coherences \citep{Schlau-Cohen2012,Westenhoff2012}.
Secondly, we Fourier transformed the complex-valued 2DES data along
$t_{2}$. This yields oscillation maps of $+\omega_{2}$ and $-\omega_{2}$
(2D slices of the 3D spectrum), with the positive and negative frequencies
separating the system response evolving as $\mathrm{e}^{-i\omega_{2}t_{2}}$
or $\mathrm{e}^{+i\omega_{2}t_{2}}$during the population time $t_{2}$,
which is very useful in separating different contributions \citep{Seibt2013,Li2013}.
When measured with the cross-polarised pulse sequence, the amplitude
maps at $\mathrm{\omega_{2}=-560\,cm^{-1}}$ and $\mathrm{\omega_{2}=+560\,cm^{-1}}$
show peaks on the off-diagonal (rephasing pulse order) and diagonal
(non-rephasing pulse order) (Fig.~\ref{fig:fig2}b). A similar pattern
is observed for the coherence at $\mathrm{\omega_{2}=\pm650\,cm^{-1}}$
(Supplementary information Fig.~S2). This pattern is as expected
for electronic coherence between the two excitonic states \citep{Butkus2012}.

Thus, the coherences at 560 and $\mathrm{650\,cm^{-1}}$ for $\mathrm{RC_{sph}}$,
if interpreted according to the current understanding of the molecular
electronic response, would be concluded to be electronic. However,
we determined that the lifetime of the two coherences is longer than
$\mathrm{1.6\,ps}$ (see Supplementary information Fig.~S5), which
is by an order of magnitude longer than the lifetime of the underlying
B and H states. This paradox prompted us to search for an alternative
photophysical explanation.

\begin{figure}[H]
\begin{centering}
\includegraphics[width=0.5\columnwidth]{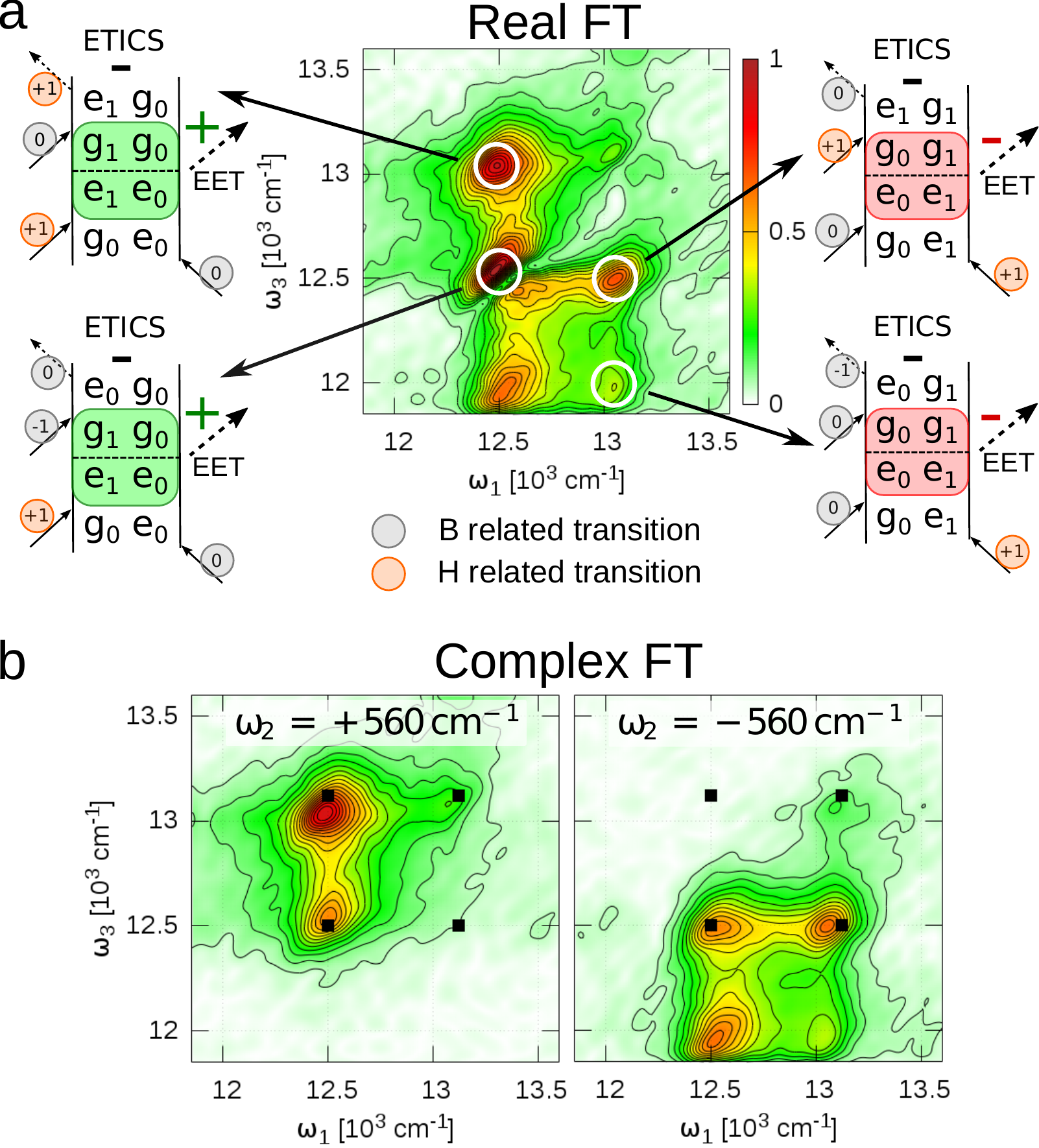}
\par\end{centering}

\caption{Cancellation of the signals at the diagonal peak $\mathrm{12500\,cm^{-1}}$
in all-parallel measurement reveals the ETICS phenomenon. (a) The
oscillation map is shown for \textbf{$\omega_{2}=560\,\protect\icm$,}
computed from real-valued rephasing spectra measured between $t_{2}\mathrm{=216-1740\,fs}$.
The associated Feynman diagrams show all the rephasing ETICS pathways,
which are complementary to their SE analogues (Supplementary information
Fig.~S3). $g$ and $e$ denote the ground and electronically excited
states, the subscript indicates the vibrational quantum number. Mixing
between $e_{1}$ of B and $e_{0}$ of H states is implied whenever
$e_{1}$ appears in the diagrams. The number next to the transition
arrows indicates the change of the vibrational quantum number. The
transition in the shaded parts of the diagrams corresponds to the
energy transfer with indicated \textbf{$\omega_{2}$} sign next to
them. The overall sign of ETICS signals is opposite to the SE analogues
and is shown above each diagram. (b) Decomposition of the oscillation
map shown in (a) to \textbf{$\pm\omega_{2}$} frequency maps by computing
the Fourier transform of the complex-valued data shows two distinct
contributions to the diagonal. Constructive interference would be
observed on the diagonal between SE \textbf{($+\omega_{2}$)} and
GSB (\textbf{$-\omega_{2}$}) signals (Supplementary information Fig.~S3).
However, the ETICS process changes the sign of the SE analogue, which
results in the observed cancellation.}

\label{fig:fig3} 
\end{figure}

\subsection*{ETICS offers a comprehensive interpretation for the observed coherences.}

To resolve this mismatch we inspect the oscillation maps at $\mathrm{\omega_{2}=560\,cm^{-1}}$,
measured with the all-parallel pulse sequence. The general pattern
of the map computed from the complex-valued data is consistent with
a combination of coherences in the excited state (stimulated emission,
SE) and ground state (ground state bleach, GSB) pathways (Fig.~\ref{fig:fig3}b,
Supplementary information, section 2). We do not consider excited
state absorption (ESA) here, because of its small contribution (see
\ref{fig:fig1}b). When computing the oscillation map from real-valued
data, a cancellation of the signal on the B peak diagonal appears
(Fig.~\ref{fig:fig3}a). This is inconsistent with the simple sum
of the SE and GSB pathways, as these pathways should interfere constructively.
However, since the data with $t_{2}<\mathrm{216\,fs},$ was omitted
in our computation of oscillation maps the SE signal are expected
to be negligible in the oscillation maps. This is because the energy
transfer to $\mathrm{P^{+}}$ is essentially complete at this time.

We therefore propose that all coherences observed in the maps (Fig.
\ref{fig:fig3}ab) are in the ground state. For the SE-like pathways, the first
two laser pulses create a coherence which is either purely vibrational
in the B excited state or it has mixed vibronic origin \citep{Christensson2012}.
Then, the coherence is shifted to the ground state. Here, vibrational
wavepacket can live for several picoseconds, independent of the excited
state lifetime. The shift of coherence is the key component of our
proposal. We consider that it is induced by the energy transfer from
B to $\mathrm{P^{+}}$ \citep{Mancal2014} and we therefore term this
process \emph{Energy Transfer Induced Coherence Shift} (ETICS) (Fig.~\ref{fig:fig4}c).

ETICS response pathways presented in Fig.~\ref{fig:fig3} show that
they should produce signals, which have an opposite sign ($\pi$-phase
shift) compared to the corresponding signals from SE pathways \citep{Mancal2014}
(Supplementary information Fig.~S3). The zero signal node in the
B diagonal (Fig.~\ref{fig:fig3}a) is then readily explained, because
the ground state bleach pathways and ETICS pathways overlap and cancel
out \footnote{The lack of total cancellation is accounted for by the opposite phase
sweep across the oscillation peaks in the GSB and ETICS signals (see
Supplementary information section 2).}. The inclusion of the ETICS pathways also agrees with the observed
relative amplitudes of the peaks in the oscillation maps in Fig.~\ref{fig:fig3}
(see Supplementary information section 2). Notably, we find the nodal
line in other oscillation frequency maps as well, which is consistent
with the ETICS process, which should occur for all vibrational wavepackets
(Supplementary information Fig. S1 and S2).

In ETICS, the SE coherence shifts to the ground state within $t_{2}\mathrm{\sim200\,fs}$,
which should result in a $\pi$-phase shift in the oscillatory signal.
This behaviour is observable at a spectral position, which is clear
of the GSB oscillation signals from impulsive Raman scattering \citep{Tiwari2013}.
Therefore, we search for the phase flip in the kinetic trace of the
upper cross-peak in the cross-polarisation measurement (Fig.~\ref{fig:fig4}ab),
where only one ETICS pathway is expected. The real and imaginary oscillating
residuals are fitted simultaneously in the time window $t_{2}=\mathrm{216-1620\,fs}$
to a sum of two complex exponentials, which account for the two frequencies
at $560\,\mathrm{cm^{-1}}$ and $650\,\mathrm{cm^{-1}}$. The fits
are then extrapolated to $t_{2}=0$. Indeed, a $\pi$-phase shift
is observed in the absorptive (real) and refractive (imaginary) part
of the signal during the characteristic time of the energy transfer
from B to $\mathrm{P^{+}}$ (see Fig.~\ref{fig:fig4}). The agreement
with the expected behaviour for ETICS is excellent, especially when
considering that the initial oscillating signal up to $t_{2}\mathrm{\leq200\,fs}$
can contain different contributions, including an electronic H-B coherence,
purely vibrational coherences, or coherences of the mixed vibronic
origin \citep{Christensson2012}.

\subsection*{Generality of ETICS.}

How general is the ETICS phenomenon? One strict requirement is that
energy transfer occurs while vibrations cool. Another requirement
is that the vibrational modes on the ground and excited electronic
states are similar. Both requirements are often fulfilled in photosynthetic
proteins, including reaction centres and light-harvesting complexes.
Thus ETICS can be expected to be general.

Importantly, ETICS retains exactly the same beating patterns as the
corresponding SE pathways. Therefore, its spectral signatures can
be easily misinterpreted as long-lived coherences with electronic
character \citep{Westenhoff2012}. For example, the oscillation maps
of a plant reaction centre protein by Fuller et al. \citep{Fuller2014}
show that the signal on the diagonal line cancels, which is not reproduced
by the model proposed in their study, but could be explained by the
ETICS process.

Moreover, the hypothesis that energy transport in photosynthetic proteins
is wave-like rests on the assignment of coherences to the superposition
of excitonic states \citep{Engel2007,Collini2010,Schlau-Cohen2012}.
ETICS provides an alternative coherence assignment and therefore calls
for reconsideration of wave-like energy transfer. Clearly, a careful
analysis of beatings in other photosynthetic proteins is needed for
accurate designation of contributing photophysical phenomena.

\subsection*{Vibronic mixing in $\mathrm{RC_{sph}}$.}

Finally, we clarify why long-lived coherences with certain frequencies
(i. e. $\mathrm{560\,cm^{-1}}$ and $\mathrm{650\,cm^{-1}}$) are
observed in the cross-polarised measurement, even though this pulse
sequence excludes all vibrational coherences generated by the Franck-Condon
excitations \citep{Schlau-Cohen2012,Westenhoff2012}. This is revealed
by the signals in the oscillation maps for $\omega_{2}=\pm560\,\mathrm{cm^{-1}}$
(Fig.~\ref{fig:fig2}b, for $\omega_{2}=\pm650\,\mathrm{cm^{-1}}$,
see Supplementary information Fig.~S2). We find that the observed
oscillation patterns for the cross-polarised pulse sequence (Fig.~\ref{fig:fig2}b)
match well with only two ETICS pathways in the rephasing signal (see
Supplementary information for detailed pathway analysis). These pathways
can only contribute to the signal if the two transitions $\mathrm{g_{0}}\rightarrow\mathrm{e_{0}}$
and $\mathrm{g_{0}}\rightarrow\mathrm{e_{1}}$ (see Feynman diagrams
in Fig.~\ref{fig:fig3}a) have non-parallel transition dipole moments.
Thus the initially excited states in the $\mathrm{RC_{sph}}$ aggregate
must be vibronically coupled, i.e. state $\mathrm{e_{1}}$ consists
of the non-adiabatic mixture of the vibronically excited state of
accessory bacteriochlorophyll and the electronically excited state
of bacteriopheophytin. This implies that vibrations are at least partly
delocalized between accessory bacteriochlorophyll and bacteriopheophytin.
Thus, we find that vibronic mixing occurs in $\mathrm{RC_{sph}}$
and that is is essential for the \textit{observation} of ETICS pathways
in the cross-polarised measurement. However, it is not necessary for
ETICS to occur.

\begin{figure}[H]
\begin{centering}
\includegraphics[width=0.4\columnwidth]{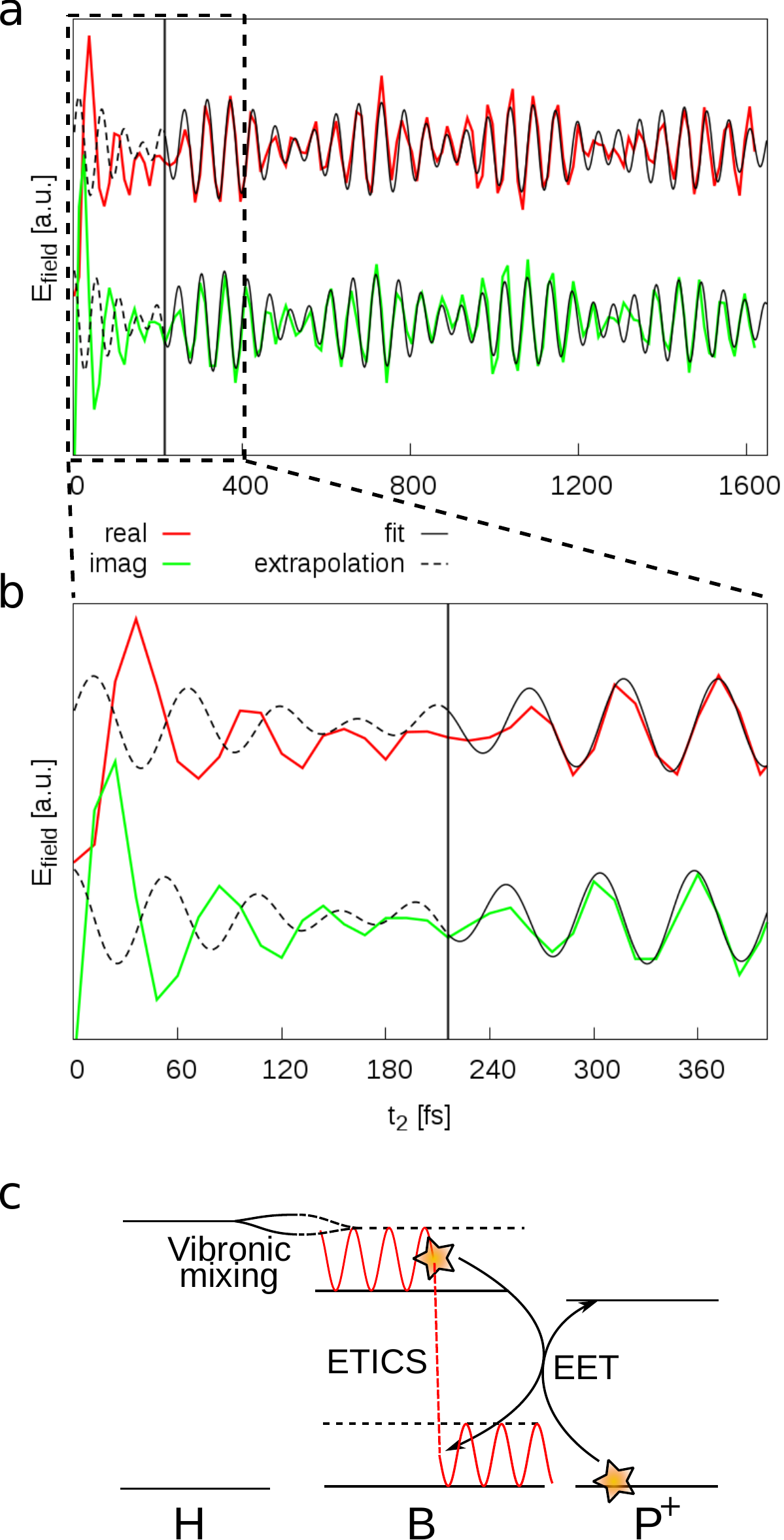} 
\par\end{centering}

\caption{The ETICS process observed in the time-domain. (a) Evolution of the
real (red) and imaginary (green) parts of the upper cross-peak in
the cross-polarised measurement and a fit by the sum of two complex
exponentials (black) for $t_{2}>\mathrm{216\,fs}$. Extrapolation
of the fit to $t_{2}=\mathrm{0\,fs}$ (dashed line) demonstrates a
$\pi$-phase change, as compared to the later population times. (b)
Zoom into the first $\mathrm{400\,fs}$ of the population time. (c)
A scheme of the ETICS process showing how initially created excited
state coherence is shifted to the ground state during the energy transfer
step to $\mathrm{P^{+}}$. Vibronic mixing between bacteriochlorophyll
and bacteriopheophytin states leads to the observable beating signals
in the cross-polarised measurement for vibrational frequencies close
to the B--H electronic gap. Observation of ETICS identifies a hot
energy transfer pathway where the access energy is damped in the donor
(b) ground state on the picosecond time scale.}

\label{fig:fig4} 
\end{figure}

Vibronic mixing has recently been suggested for natural \citep{Womick2011,Christensson2012,Chin2013,Tiwari2013}
and synthetic \citep{Lim2015} light-harvesting complexes. It has
been proposed to facilitate energy transfer \citep{Womick2011,Tiwari2013,Perlik2015},
and it has also been used to explain long-lived quantum coherences
\citep{Christensson2012,Tiwari2013,Ferretti2013,Fuller2014,Romero2014,Ryu2014}.
In particular, Christenson et al. predicted that mixing of atomic
and electronic degrees of freedom leads to the slow dephasing of mixed
vibronic coherences in the excited state\citep{Christensson2012}.
However, because the lifetime of the excited states is shorter than
200 fs, we do not consider this hypothesis a viable explanation for
the long-lived beatings in $\mathrm{RC_{sph}}$.

Tiwari et al. proposed that vibronic coupling opens up a new response
pathway by which direct excitation of mixed ground state coherences
is facilitated \citep{Tiwari2013}. This model was used to interpret
the long-lived oscillating signals observed in the two-colour photon
echo experiments of $\mathrm{RC_{sph}}$ \citep{Ryu2014}. However,
we do not find the predicted strong asymmetry of oscillation amplitudes
in the cross-peaks above and and below the diagonal in the 2D spectra
(see Fig.~\ref{fig:fig2}b and Supplementary information section
4 for details). Moreover, the pathways suggested in Ref.~\citep{Tiwari2013}
are inconsistent with the phase shift which we observe on the 200-fs
time scale (Fig.~\ref{fig:fig4}) leading to cancellation of the
oscillating signals appearing on the diagonal (Fig.~\ref{fig:fig3}a).
Thus, we conclude that direct excitation of ground state vibrational
coherence as proposed by Tiwari et al. is not dominant in $\mathrm{RC_{sph}}$.

\subsection*{Hot energy transfer in $\mathrm{RC_{sph}}$ facilitated by vibrational
dissipation in the ground state.}

Although we identify ETICS as a shift of vibrational coherences between
electronic states, equivalent shifts of populations are equally possible.
This indicates energy transfer from hot B to the special pair, while
a vibrational quantum is left on the accessory bacteriochlorophyll.
Thus, the vibrational modes on the accessory bacteriochlorophylls
work as a sink by taking up the excess energy from energy transfer
between B and the special pair. In this study we show that many vibrational
modes on accessory bacteriochlorophylls can acts as an energy sink
for the energy transfer step \citep{Schulze2016}. Energy conversion demands the existence
of such modes, but it has not been possible to detect them spectroscopically.
Consequently, these modes have usually been approximated as spectral
densities in theory of energy transfer. Our finding opens a new possibility
to directly follow the dissipation of excess energy in excitation
transfer into vibrations of the molecules.

We identified the special modes which have frequencies close to the
B-H resonance. They have a twofold effect on energy transfer. Firstly,
vibronic mixing increases the rate of energy transfer between B and
H \citep{Womick2011,Tiwari2013}, leading to the efficient population
of hot B. Secondly, these modes accelerate energy transfer further
from hot B to the special pair, before vibrational relaxation on B
takes place on a picosecond time scale. As seen from a molecular perspective,
the atoms of the chromophore are already set to oscillate within an
appropriate mode in the excited state, priming the molecules for energy
transfer. Thus certain modes appear to be an integral part of downhill
energy transfer, increasing its overall efficiency. The availability
of these special modes of chromophores may explain the extraordinarily
fast energy transfer rates in reaction centre proteins. Also, the
possibility that the protein uses these modes to control the directionality
of the flow of energy should not be disregarded.

\section*{Materials and Methods}

\label{sec:mat}

$\mathrm{RC_{sph}}$ were isolated from $Rhodobacter\,sphaeroides$,
R-26 following standard procedures described elsewhere \citep{Farhoosh1997}
with modifications described in Supporting Information. To oxidise
$\mathrm{RC_{sph}}$, potassium ferricyanide ($\mathrm{K_{3}Fe(CN)_{6}}$)
was added to a final concentration of 150\,mM. Samples were mixed
with glycerol at 35:65 (v/v) and cooled down to $\mathrm{77\,K}$
in a $\mathrm{0.5\,mm}$ cell made of fused silica. The samples typically
had an optical density of 0.2 to 0.3 at $\mathrm{800\,nm}$.

Spectroscopy setup: The data acquisition protocol and the analysis
were described previously \citep{Brixner2004,Augulis2011,Augulis2013}.
Briefly, a noncolinear optical amplifier was pumped by the $\mathrm{1030\,nm}$
fundamental of a Pharos laser system (Light Conversion Ltd). The resulting
$\sim\mathrm{17\,fs}$ laser pulses were split into four beams using
a beamsplitter and a transmissive diffraction grating. Spherical optics
were used to focus all beams to $\sim\SI{100}{\micro\metre}$ on the
sample spot. The forth beam was attenuated by and OD=2 filter. The
first two beams were simultaneously chopped by mechanical choppers
and a double frequency lock-in detection scheme was used. Interferograms
were continuously detected by the CCD camera (PIXIS, Princeton Instruments).

A repetition rate of $\mathrm{20\,kHz}$ with excitation energies
of $\mathrm{2\,nJ}$ and $\mathrm{4\,nJ}$ per pulse were used for
all-paralell and cross-polarised measurements, respectively. This
translates to excitation density of $\mathrm{10^{14}}$ photons per
pulse per $\mathrm{cm^{2}}$ for the $\mathrm{2\,nJ}$ pulse energy.
The population time step was $\mathrm{12\,fs}$, which defines the
high frequency cut-off of frequency to $\omega_{2}=1380\,\mathrm{cm^{-1}}$.
The coherence time $t_{1}$ was scanned between $-199.5$--$294\,\mathrm{fs}$
($-171.5$--$273\,\mathrm{fs}$) with the $\mathrm{1.75\,fs}$ step
for all-parallel (cross-polarised) measurements, respectively. In
both axis, the resolution was typically $\mathrm{50\,cm^{-1}}$ ($\mathrm{58\,cm^{-1}}$)
for the all-parallel (cross-polarised) experiments (Supplementary
information, section 6).

The sequences of 2D spectra in $t_{2}$ were Fourier transformed to
generate three-dimensional FT spectra. To reduce the complexity of
the third order response signals and avoid multiple resonant and nonresonant
contributions during the pulse overlap, we analysed the oscillating
signals in a following way. We extracted the oscillatory components
in $t_{2}$ by subtracting multi-exponential fits with complex amplitude
prefactors from each ($\omega_{1},\omega_{3}$) data point. The Fourier
transforms of the remaining oscillating residuals in $t_{2}$ yields
the three-dimensional FT spectra. By slicing these spectra in $\omega_{2}$
we extracted the ($\omega_{1},\omega_{3}$) coherence amplitude and/or
phase maps of oscillations with $\omega_{2}$ frequency. The resolution
along $\omega_{2}$ is determined by the length of the $t_{2}$ scan,
$0-1740\mathrm{\,fs}$ and $0-1620\,\mathrm{fs}$ for all-paralell
and cross-polarised, respectively and was $\mathrm{22\,cm^{-1}}$
and $\mathrm{24\,cm^{-1}}$ for all-paralell and cross-polarised,
respectively.

\section*{Acknowledgements}

We would like to thank Darius Abramavicius, Vytautas Butkus, Leonas
Valkunas, Tõnu Pullerits and Tomáš Mančal for helpful discussions.
The work in Lund was supported by the Swedish Research Council and
Knut and Alice Wallenberg foundations. S.W. and P. E. acknowledge
funding from the Foundation of Swedish Research Sweden, the Olle Engkvist
Byggmästare Foundation and the Swedish Research Council.

\section*{Author contributions}

D.P., S. W. and D.Z. conceived and designed the experiments. D.P.
performed the experiments. D.P. analysed the data. P.E. grew and purified
the sample. D.P., D.Z., SW and E.P. wrote the paper. All authors discussed
the results and commented on the manuscript.

\section*{Additional information}

The authors declare no competing financial interests.

 \bibliographystyle{naturemag}
\bibliography{references}

\end{document}